\documentclass[10pt, letterpaper, conference]{IEEEtran}
\IEEEoverridecommandlockouts

\usepackage{cite}
\usepackage{amsmath,amssymb,amsfonts}
\usepackage{bm}
\usepackage{algorithmic}
\usepackage{graphicx}
\usepackage{textcomp}
\usepackage{xcolor}
\usepackage{subcaption}
\usepackage{algorithm}
\usepackage{comment}
\usepackage{balance}

\usepackage[
    bookmarks=false,
    draft
]{hyperref}
\def\BibTeX{{\rm B\kern-.05em{\sc i\kern-.025em b}\kern-.08em
    T\kern-.1667em\lower.7ex\hbox{E}\kern-.125emX}}

\def\Htran{\mbox{\tiny $\mathrm{H}$}}
\begin{document}
\setlength{\columnsep}{0.24in}
\setlength{\textfloatsep}{8pt plus 2pt minus 2pt}
\setlength{\floatsep}{8pt plus 2pt minus 2pt}
\title{Movable Antenna Arrays with Imperfect Channel State Information in Rich Scattering Environments}

\author{Yizhen~Zhao, Amna~Irshad and Emil~Bj{\"o}rnson \\
\IEEEauthorblockA{\textit{Department of Communication Systems, KTH Royal Institute of Technology, Stockholm, Sweden}\\  Email: \{yizhen,amnai,emilbjo\}@kth.se}
\thanks{The research was supported by the Grant 2022-04222 from the Swedish Research Council.}

}

\maketitle

\begin{abstract} 
The growing demand for high spectral efficiency in 6G and beyond has driven research into adaptive antenna architectures capable of exploiting the spatial structure of multipath propagation channels. Conventional base station arrays are deployed with fixed element positions and cannot adapt to the instantaneous spatial structure of the propagation channel. In contrast, movable antenna (MA) systems enable dynamic reconfiguration of antenna positions, allowing the array geometry to track the channel characteristics of the current user set. While prior MA studies have demonstrated significant gains under perfect or statistical channel state information (CSI) assumptions, the interplay between imperfect instantaneous CSI, antenna placement optimization, and precoder design has received limited attention. This paper addresses this gap by proposing a practical end-to-end framework encompassing uplink pilot transmission, MMSE channel estimation under a clustered multipath model, and downlink ZF precoding designed from estimated CSI. Antenna positions are optimized via particle swarm optimization under two objectives: sum-rate maximization and max-min fairness. Simulation results show that MA gains are most pronounced under sparse, near-LoS propagation and diminish as channel richness increases. Furthermore, we reveal a fundamental coupling between array geometry and precoder design: a fairness-oriented antenna geometry encodes spatial fairness information that is only recoverable when evaluated with a compatible power allocation strategy. A mismatched precoder can completely mask the geometric advantage, leading to misleading conclusions about the robustness of antenna placement to the choice of optimization objective. These findings provide practically relevant guidance for the design and evaluation of movable antenna systems under realistic operating conditions.
\end{abstract}

\begin{IEEEkeywords}
Movable antennas, MMSE channel estimation, particle swarm optimization, multi-user MIMO, clustered channel model.
\end{IEEEkeywords}

\section{Introduction}

Massive MIMO has become a cornerstone of 5G wireless systems, offering substantial improvements in spectral efficiency and spatial multiplexing capability~\cite{Marzetta2010a,Bjornson2019c}. Extending these gains to beyond-5G and 6G scenarios typically requires scaling up the number of antennas, which significantly increases hardware cost and system complexity. This challenge motivates the investigation of flexible antenna architectures capable of achieving high performance with a reduced number of elements. Nevertheless, most existing systems employ antenna elements fixed at predetermined positions on the array aperture, typically with half-wavelength spacing between elements. While such fixed geometries are simple to deploy and analyze, they are fundamentally constrained because, once deployed, the array cannot adapt to the actual user distribution or propagation environment, both of which are time-varying.

Recent years have seen growing interest in \emph{movable antenna} (MA) systems \cite{Bjornson2019c}, also referred to as fluid antennas~\cite{New2026} or mechanically reconfigurable arrays, in which individual antenna elements can reposition within a bounded deployment region~\cite{Zhu2025a,movgeneral}. By adjusting antenna positions, the BS can reconfigure its structure to better align with the dominant directions of signal propagation, more efficiently exploit spatial degrees of freedom, and reduce spatial correlation among user channels. These properties make MA systems particularly attractive for multi-user downlink scenarios, 
where inter-user interference suppression depends critically on the spatial separability of user channels.

Motivated by these advantages, substantial research effort has been directed toward MA-aided communication systems. The capacity benefits of MA systems in single-user scenarios were characterized in~\cite{movgeneral}, showing that antenna position optimization can achieve spatial multiplexing gains approaching theoretical bounds. In multi-user systems, optimizing MA positions has been shown to improve achievable rates~\cite{Zhu2025a,PIA2025}, suppress inter-user interference, and enable flexible beamforming. Furthermore, position optimization based on statistical CSI~\cite{MA26} has been studied to reduce dependence on instantaneous channel knowledge. Despite these promising results, most existing works assume either perfect instantaneous CSI or ideal statistical CSI at the BS, which do not reflect the conditions encountered in practical deployments.

In practice, the BS must estimate user channels from uplink pilot signals, and this estimation is inherently imperfect due to noise and finite pilot length. The resulting errors degrade both the precoding quality and the reliability of any utility function used to drive antenna placement. Optimizing ergodic rates provides a principled way to handle this uncertainty, as they average over channel estimation errors across coherence intervals. While pilot-based estimation for MA systems has been studied under field-response channel models~\cite{Zhang_2025,Wenyan23,Xiao2024,Ruoyu2024}, these works focus on reconstructing the spatial channel function across the aperture rather than quantifying how estimation error propagates through the precoder and into the placement objective. To the best of our knowledge, no prior work has evaluated the complete chain of uplink pilot transmission, MMSE channel estimation, ZF precoding, and PSO-based antenna placement within a unified framework that explicitly accounts for imperfect CSI throughout.

\subsection*{Contributions}
The main contributions of this paper are as follows:
\begin{itemize}
    \item We propose a complete end-to-end framework for MA-assisted multi-user downlink systems that incorporates uplink MMSE channel estimation, precoding, and PSO-based antenna placement optimization, using ergodic rate expressions that capture imperfect CSI.
    \item We show that the performance gains of movable antennas are highly dependent on channel structure: significant improvements are obtained in sparse, highly directional channels, while the advantage diminishes in rich scattering environments.
    \item We reveal a fundamental coupling between array geometry and precoder design: a fairness-oriented MA geometry only yields min-rate gains when evaluated with a compatible power allocation strategy, and a mismatched precoder can completely suppress the geometric advantage.

\end{itemize}

The remainder of this paper is organized as follows. Section~\ref{sec:system} presents the system model, including the channel model and pilot-based channel estimation. Section~\ref{sec:problem} formulates the antenna position optimization problem. Section~\ref{sec:solution} describes the proposed solution approach. Simulation results and performance analysis are provided in Section~\ref{sec:results}, followed by conclusions in Section~\ref{sec:conclude}.

\section{System Model}\label{sec:system}

\begin{figure}[t]
\centering
{\includegraphics[width=0.9\columnwidth]{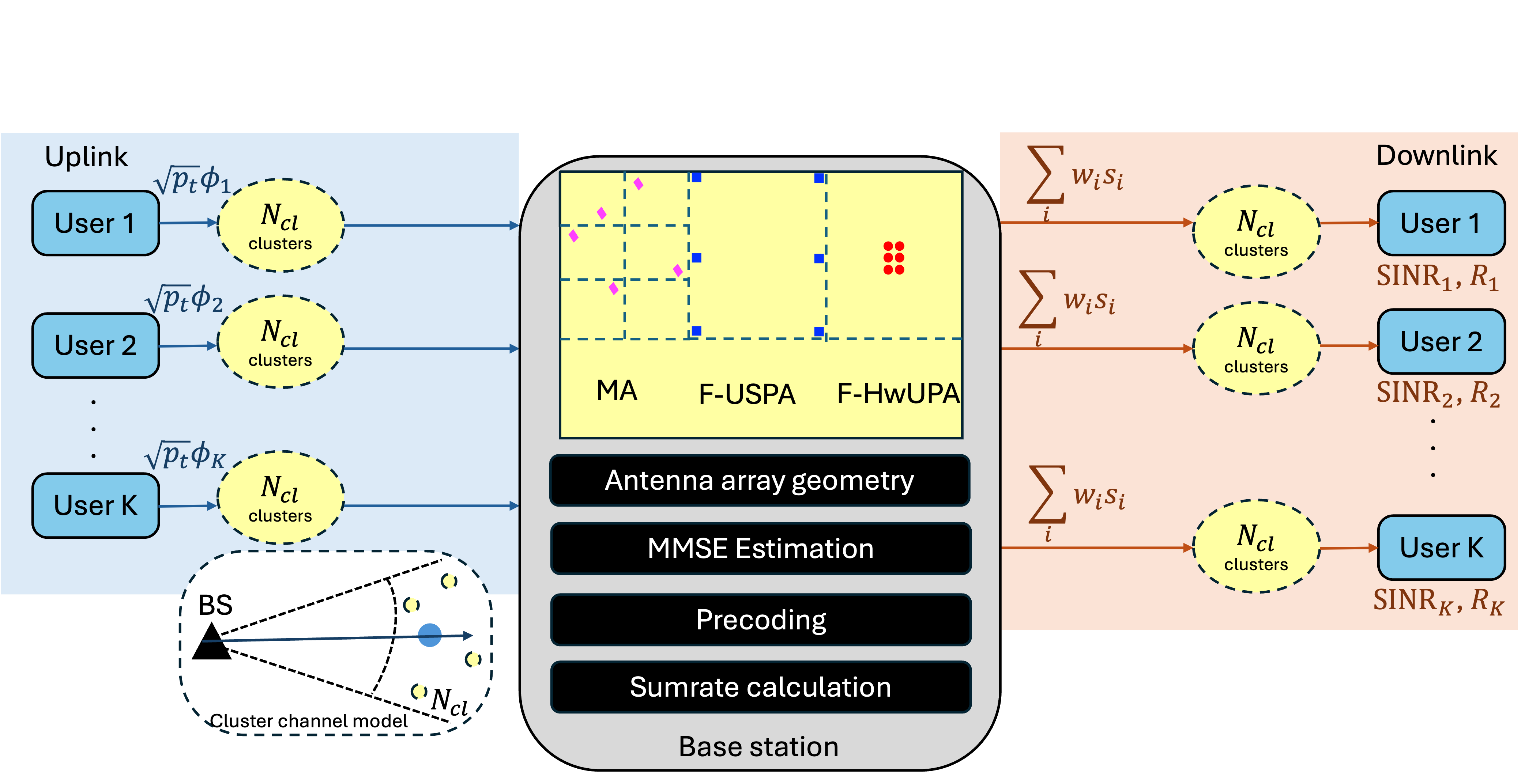}}
\caption{Block diagram of the system where a base station serves $K$ users having clustered multipath channels.}\label{fig:system}
\end{figure}

We consider a multi-user downlink system where a base station (BS) equipped with $N$ MAs serves $K$ single-antenna users. The propagation channel between the BS and $k$-th user is modeled using a clustered channel representation~\cite{Saleh1987}. More precisely, each user channel consists of $N_{\mathrm{cl}}$ scattering clusters whose directions are distributed around the line-of-sight (LoS) direction with a limited angular spread as shown in Fig.~\ref{fig:system}. We consider time-division duplexing, so the channels are identical in uplink and downlink.

\subsection{Clustered Channel Model}
Let $\phi_{k,i}$ and $\theta_{k,i}$ denote the azimuth and elevation angles associated with the $i$-th cluster of the $k$-th user. The corresponding wave vector is defined as
\begin{equation}
\mathbf{k}_{k,i} =\frac{2\pi}{\lambda}
\begin{bmatrix}
\cos\theta_{k,i} \cos\phi_{k,i} \\
\cos\theta_{k,i} \sin\phi_{k,i} \\
\sin\theta_{k,i}
\end{bmatrix}.
\end{equation}
If the $n$-th antenna element is located at the position $\mathbf{t}_n \in \mathbb{R}^3$, the array response vector for the $i$-th cluster becomes
\begin{equation}
\mathbf{a}(\phi_{k,i},\theta_{k,i}) =
\left[
e^{j\mathbf{k}_{k,i}^{\mathrm{T}} \mathbf{t}_1},
\ldots,
e^{j\mathbf{k}_{k,i}^{\mathrm{T}} \mathbf{t}_N}
\right]^{\mathrm{T}},
\end{equation}
where $\lambda$ is the carrier wavelength. The channel vector of the $k$-th user is modeled as
\begin{equation}
\mathbf{h}_k = \sum_{i=1}^{N_{\mathrm{cl}}} c_{k,i} \,\mathbf{a}(\phi_{k,i},\theta_{k,i}),
\end{equation}
where $c_{k,i} \sim \mathcal{CN}(0,\beta_{k,i})$ is the complex fading coefficient of the $i$-th cluster and $\beta_i$ is the cluster power \cite{molisch2011wireless}. The corresponding spatial covariance matrix is
\begin{equation}
\mathbf{R}_k = \mathbb{E} \{ \mathbf{h}_k\mathbf{h}_k^{\Htran} \} =\sum_{i=1}^{N_{\mathrm{cl}}} \beta_{k,i}\,
\mathbf{a}(\phi_{k,i},\theta_{k,i})\mathbf{a}^{\Htran}(\phi_{k,i},\theta_{k,i}).
\end{equation}

\subsection{Pilot-Based MMSE Channel Estimation}

During the uplink training phase, each user transmits an orthogonal pilot sequence of length $\tau_p \geq K$. The received pilot matrix at the BS is
\begin{equation}
\mathbf{Y} = \sqrt{p_t}\,\mathbf{H}\mathbf{\Phi}^T + \mathbf{N},
\end{equation}
where $p_t$ is the common pilot transmit power, $\mathbf{H} = [\mathbf{h}_1,\ldots,\mathbf{h}_K]$, $\mathbf{\Phi} \in \mathbb{C}^{\tau_p \times K}$ is the pilot matrix satisfying $\mathbf{\Phi}^H\mathbf{\Phi} = \tau_p \mathbf{I}_K$, and $\mathbf{N}$ is the additive complex Gaussian noise with independent $\mathcal{CN}(\mathbf{0}, N_0\mathbf{I})$-entries, where $N_0$ is the noise variance.
The BS obtains MMSE channel estimates from uplink pilots. The estimate of $\mathbf{h}_k$ is given by~\cite{Bjornson2017}
\begin{equation}
\hat{\mathbf{h}}_k = \sqrt{p_t\tau_p}\,\mathbf{R}_k\,\mathbf{\Psi}_k^{-1}\mathbf{z}_k,
\end{equation}
where $\mathbf{\Psi}_k = p_t\tau_p\,\mathbf{R}_k + N_0\mathbf{I}$. The channel estimation error $\tilde{\mathbf{h}}_k = \mathbf{h}_k - \hat{\mathbf{h}}_k$ is independent of $\hat{\mathbf{h}}_k$ by the orthogonality principle and Gaussian distribution, and its covariance matrix is
\begin{equation}
\mathbf{C}_k = \mathbf{R}_k - p_t\tau_p\,\mathbf{R}_k\,\mathbf{\Psi}_k^{-1}\mathbf{R}_k.
\end{equation}

\subsection{Downlink Precoding and Achievable Ergodic Sum Rate}
The BS applies linear precoding based on the estimated channels, where the precoding vectors  $\mathbf{W}=[\mathbf{w}_1, \mathbf{w}_2, \cdots,\mathbf{w}_K]$ are designed to suppress inter-user interference based on the estimated channel matrix. 
Power is allocated across users by water-filling over the effective channel gains $|\hat{\mathbf{h}}_k^{\Htran}\mathbf{w}_k|^2$.
Due to the imperfect channel estimate, the received signal at $k$-th user contains residual interference from the estimation error. The resulting downlink received signal is
\begin{equation}
y_k = \hat{\mathbf{h}}_k^{\Htran} \mathbf{w}_k s_k +
\sum_{\substack{i=1\\i\neq k}}^{K} \hat{\mathbf{h}}_k^{\Htran} \mathbf{w}_i s_i +
\sum_{i=1}^{K} \tilde{\mathbf{h}}_k^{\Htran} \mathbf{w}_i s_i + n_k,
\end{equation}
where $s_k \sim \mathcal{CN}(0, P_k)$ is the transmitted data symbol and $n_k \sim \mathcal{CN}(0,\sigma^2)$ is receiver noise. 
We assume that the UEs also know the imperfect channel estimates, which implies that the received signal contains residual interference caused by imperfect CSI. By treating the interference and estimation errors as worst-case Gaussian noise, a standard ergodic achievable rate can be obtained~\cite{Bjornson2017}. For $k$-th user, the ergodic achievable rate is
\begin{equation}
R_{\mathrm{k}} =\mathbb{E}\left\{ \log_2(1 + \mathrm{SINR}_k)\right\} ,
\label{eq:sumrate}
\end{equation}
where the effective instanteneous SINR is 

\begin{equation}\label{eq:sinr}
\mathrm{SINR}_k =
\frac{P_k\,|\hat{\mathbf{h}}_k^{\Htran}\mathbf{w}_k|^2}{\sum_{i \neq k}P_i\,|\hat{\mathbf{h}}_k^{\Htran}\mathbf{w}_i|^2
+ \sum_{i=1}^{K} \mathbf{w}_i^{\Htran} \mathbf{C}_k \mathbf{w}_i + \sigma^2}.
\end{equation}

The SINR's denominator explicitly includes the term $\sum_{i=1}^{K}\mathbf{w}_i^{\Htran}\mathbf{C}_k\mathbf{w}_i$, which captures the aggregate interference power arising from channel estimation errors. This term is absent in ideal CSI analysis, and its inclusion is essential for an accurate evaluation of system performance under practical conditions.

\section{Problem Formulation}\label{sec:problem}
The objective of this paper is to optimize the positions of the $N$ BS antenna elements to maximize a general network utility function of the user rates. Specifically, we consider
\begin{align}
\underset{\mathbf{t}_1,\ldots,\mathbf{t}_N}{\mathrm{maximize}} \quad 
& U\big(R_1,\ldots,R_K\big) \label{eq:opt}\\
\mathrm{subject~to} \quad
& \mathbf{t}_n \in \mathcal{C}_n, \quad \forall\, n, \label{eq:c1}\\
& \|\mathbf{t}_i - \mathbf{t}_j\| \geq \frac{\lambda}{2}, \quad \forall\, i \neq j. \label{eq:c2}
\end{align}
Constraint~\eqref{eq:c1} restricts each antenna element to move within a corresponding bounded feasible region $\mathcal{C}_n$, representing mechanical deployment constraints. Constraint~\eqref{eq:c2} enforces a minimum inter-element spacing of $\lambda/2$ to limit mutual coupling effects~\cite{hong2025fluidantennaempowering5g}, and $R_k$ is given in~\eqref{eq:sumrate}.

The rate expression in~\eqref{eq:sumrate} explicitly captures the impact of imperfect CSI through the estimation error covariance $\mathbf{C}_k$, making the optimization fundamentally different from ideal-CSI formulations. The utility function $U(\cdot)$ is assumed to be monotonically increasing in each argument and can represent different design objectives, such as:
\begin{itemize}
\item Sum-rate maximization: $U = \sum_{k=1}^K R_k$
\item Max-min fairness: $U = \min_k R_k$
\end{itemize}

This general formulation enables a unified treatment of different performance objectives without modifying the underlying optimization framework. The problem in~\eqref{eq:opt}--\eqref{eq:c2} is non-convex due to the nonlinear dependence of $R_{\mathrm{sum}}$ on antenna positions through the steering vectors, and the combinatorial nature of the minimum-spacing constraint. Moreover, the objective function evaluated at each candidate geometry requires a full channel estimation and precoding computation, making gradient-based methods impractical. These characteristics motivate the use of a population-based algorithm, as described in the following section.

\section{Proposed PSO-Based Solution}\label{sec:solution}
To solve the antenna placement optimization problem, we use particle swarm optimization (PSO)~\cite{PSO15}, a population-based algorithm that is well-suited to non-convex, high-dimensional, and constrained optimization problems. PSO requires no gradient information and can escape local optima through the cooperative interaction of a population of candidate solutions.

\subsection{PSO Formulation}
Each particle $p$ in the swarm represents a candidate antenna configuration $\mathbf{T}^{(p)} = [\mathbf{t}_1^{(p)}, \ldots, \mathbf{t}_N^{(p)}]$. The fitness of each particle is a general utility function $U(R_1,\ldots,R_K)$ evaluated using the achievable rates in~\eqref{eq:sumrate}. This allows the same framework to optimize different objectives, such as sum-rate maximization and max-min fairness. This computation is done using the estimated channels $\{\hat{\mathbf{h}}_k\}$ obtained from MMSE estimation. This is a critical design choice: by using estimated rather than true channels to evaluate fitness, the optimization is performed under realistic imperfect CSI conditions, consistent with what is available at the BS in practice~\cite{bjornson2024introduction}.
At each iteration $t$, the velocity and position of particle $p$ are updated as
\begin{align}
\mathbf{v}^{(p)}_{t+1} &= w\,\mathbf{v}^{(p)}_t + c_1 r_1 (\mathbf{T}^{(p)}_{\mathrm{best}} - \mathbf{T}^{(p)}_t) + c_2 r_2 (\mathbf{T}^{*}_{\mathrm{best}} - \mathbf{T}^{(p)}_t),\\
\mathbf{T}^{(p)}_{t+1} &= \mathbf{T}^{(p)}_t + \mathbf{v}^{(p)}_{t+1},
\end{align}
where $w$ is the inertia weight, $c_1$ and $c_2$ are cognitive and social acceleration coefficients, $r_1, r_2 \sim \mathcal{U}(0,1)$ are random scalings, $\mathbf{T}^{(p)}_{\mathrm{best}}$ is the personal best position of particle $p$, and $\mathbf{T}^{*}_{\mathrm{best}}$ is the global best position found by the swarm.

\subsection{Constraint Handling}
The minimum inter-element spacing constraint~\eqref{eq:c2} is incorporated into the PSO objective via a penalty term. The modified fitness function evaluated by each particle is
\begin{equation}
f_{\mathrm{pen}}\!\left(\mathbf{T}^{(p)}\right) =
U\!\left(R_1,\ldots,R_K\right)
- \alpha \cdot g\!\left(\mathbf{T}^{(p)}\right),
\label{eq:fpen}
\end{equation}
where
\begin{equation}
g\!\left(\mathbf{T}^{(p)}\right) =
\max_{i \neq j}\!
\left[\frac{\lambda}{2} - \left\|\mathbf{t}_i^{(p)} - \mathbf{t}_j^{(p)}\right\|\right]^{\!+}
\end{equation}
is the maximum spacing constraint violation across all antenna pairs, $[\cdot]^+ = \max(0,\cdot)$ denotes the positive part, and $\alpha > 0$ is the penalty weight, set to $\alpha = 1$ in all experiments. Particles violating the spacing constraint incur a fitness reduction proportional to the degree of violation, guiding the swarm toward feasible configurations over successive iterations. The box constraints in~\eqref{eq:c1} are enforced exactly through variable bounds in the PSO solver, ensuring each antenna element always remains within its feasible region $\mathcal{C}_n$.

\subsection{PSO Algorithm}
The complete PSO-based antenna placement procedure is summarized in Algorithm~\ref{alg:pso}. Steps~3--5 highlight the key feature of this framework: the MMSE estimation and ZF precoding pipeline is embedded inside the fitness evaluation loop, so that every candidate antenna geometry is assessed using the same realistic estimated-channel based utility function that the deployed system would achieve. In the simulation results, the PSO is configured with $N_p = 120$ particles, $I = 180$ iterations, inertia weight $w = 0.5$, coefficients $c_1 = 1.2$ and $c_2 = 2$, and penalty coefficient $\alpha = 5$; the algorithm terminates after $I$ iterations.

\begin{algorithm}[t!]
\small
\caption{PSO-Based MA Placement under Imperfect CSI}
\label{alg:pso}
\begin{algorithmic}[1]
\REQUIRE User positions, cluster directions $\{\mathbf{k}_i^{(k)}\}$, system params
\ENSURE Optimized antenna layout $\mathbf{T}^*$
\STATE \textbf{Initialize} swarm: draw $N_p$ particles $\mathbf{T}^{(p)}$ uniformly from $\mathcal{C}_n$; initialize velocities $\mathbf{v}^{(p)} = \mathbf{0}$; set $\mathbf{T}^{(p)}_{\mathrm{pbest}} \leftarrow \mathbf{T}^{(p)}$
\FOR{each particle $p$}
    \STATE Build array $\mathbf{A}$ from positions $\mathbf{T}^{(p)}$
    \STATE Compute $\mathbf{h}_k$, $\mathbf{R}_k$ via clustered channel model
    \STATE Run MMSE estimation $\rightarrow$ $\hat{\mathbf{h}}_k$, $\mathbf{C}_k$
    \STATE Run ZF precoding + water-filling $\rightarrow$ $\mathbf{W}$
    \STATE Compute user rates $\{R_k\}$ and evaluate $U(R_1,\ldots,R_K)(\hat{\mathbf{h}}_k, \mathbf{C}_k,\mathbf{W})$ via~\eqref{eq:sumrate}
    \STATE Evaluate $f_{\mathrm{pen}}(\mathbf{T}^{(p)})$ via~\eqref{eq:fpen}
\ENDFOR
\STATE Set $\mathbf{T}^{*}_{\mathrm{gbest}} \leftarrow \arg\max_p\, f_{\mathrm{pen}}(\mathbf{T}^{(p)})$
\FOR{$t = 1$ to $I$}
    \FOR{each particle $p$}
        \STATE Update $\mathbf{v}^{(p)}$ and $\mathbf{T}^{(p)}$ via PSO velocity equations
        \STATE Clip $\mathbf{T}^{(p)}$ to box bounds $\mathcal{C}_n$ (enforce~\eqref{eq:c1})
       \STATE Evaluate $f_{\mathrm{pen}}(\mathbf{T}^{(p)})$ (Steps 3--8 above)
        \IF{$f_{\mathrm{pen}}(\mathbf{T}^{(p)}) > f_{\mathrm{pen}}(\mathbf{T}^{(p)}_{\mathrm{pbest}})$}
            \STATE $\mathbf{T}^{(p)}_{\mathrm{pbest}} \leftarrow \mathbf{T}^{(p)}$
        \ENDIF
        \IF{$f_{\mathrm{pen}}(\mathbf{T}^{(p)}) > f_{\mathrm{pen}}(\mathbf{T}^{*}_{\mathrm{gbest}})$}
            \STATE $\mathbf{T}^{*}_{\mathrm{gbest}} \leftarrow \mathbf{T}^{(p)}$
        \ENDIF
    \ENDFOR
\ENDFOR

\RETURN $\mathbf{T}^* = \mathbf{T}^{*}_{\mathrm{gbest}}$
\end{algorithmic}
\end{algorithm}

\subsection{Computational Complexity}
The dominant cost per PSO fitness evaluation is MMSE channel estimation, requiring inversion of an $M\times M$ matrix per user at $\mathcal{O}(KM^3)$, and ZF null-space computation at $\mathcal{O}(K^2M^2)$. Since $M \geq K$, the former dominates. Averaged over $N_{\mathrm{MC}}$ realizations and across the full PSO loop of $N_p$ particles and $I$ iterations, the total complexity per scenario is $\mathcal{O}\!\left(N_p\, I\,N_{\mathrm{MC}}\left(KM^3 + K^2M^2\right)\right)$, which evaluates to approximately $7\times 10^{10}$ operations for the parameter values in Table~\ref{tab:sim_parameters}. The 
optimization is thus suited to offline deployment, where it is re-run only when large-scale channel statistics change.
\section{Numerical Results}\label{sec:results}

We evaluate the system performance with the parameter settings summarized in Table~\ref{tab:sim_parameters}. Monte Carlo simulations are performed over 100 user drops. In each realization, 50 independent channel realizations ($N_{\mathrm{MC}}$) are considered, and $K = 10$ users are dropped uniformly within an annular sector at azimuth angles in $[-60^\circ, 60^\circ]$ and distances between 10\,m and 50\,m from the BS \cite{3GPP25996,Yu2002}. The cluster directions for each realization are generated once and shared across all antenna geometry evaluations, ensuring a fair comparison. Linear precoding is employed at the BS, with ZF precoding based on the estimated channel matrix to suppress inter-user interference to the extent possible with imperfect CSI. Each antenna moves within a dedicated $5\lambda \times 5\lambda$ cell, 
spanning a total aperture of $20\lambda \times 20\lambda$.

\begin{table}[t]
\centering
\caption{Simulation Parameters}\label{tab:sim_parameters}
\renewcommand{\arraystretch}{1.2}
\begin{tabular}{lc}
\hline
\textbf{Parameter} & \textbf{Value} \\
\hline
Carrier frequency $f_c$ & 3\,GHz \\
Wavelength $\lambda$ & 0.1\,m \\
Number of BS antennas $N$ & 16 ($4\times 4$) \\
Number of users $K$ & 10 \\
Number of clusters $N_{\mathrm{cl}}$ & 5 (varied: 1, 5, 10) \\
Pilot length $\tau_p$ & $K$ \\
Pilot transmit power $p_t$ & 5 \\
Downlink SNR & 10\,dB \\
Noise power $N_0$ & 1 (normalized) \\
Downlink power $P$ & $\mathrm{SNR}_{\mathrm{linear}} \times N_0$ \\
BS height & $20\lambda$ \\
\hline
\end{tabular} 
\end{table}

To isolate the contributions of antenna geometry and CSI quality independently, each antenna configuration is evaluated under two channel conditions:
\begin{itemize}
   \item \textbf{Estimated channel} ($\hat{\mathbf{h}}_k$, \textbf{solid lines}): The ZF precoder is designed from the MMSE estimate, and the SINR in~\eqref{eq:sinr} includes $\mathbf{C}_k$. This is the realistic operating condition.
   \item \textbf{True channel / genie-aided} ($\mathbf{h}_k$, \textbf{dashed lines}): The ZF precoder is designed from the true channel with $\mathbf{C}_k = \mathbf{0}$. This is the perfect CSI upper bound for each geometry where $\hat{\mathbf{h}}_k=\mathbf{h}_k$.
\end{itemize}
The vertical gap between a scheme's dashed and solid CDF curves directly quantifies the performance loss due to imperfect CSI for that geometry. We consider two MA configurations depending on the optimization objective: sum-rate (MA-SR) and max–min fairness (MA-MM). In addition to MA, traditional fixed BS schemes are also considered and are:

\begin{itemize}
    \item \textbf{Sparse Uniform Planar Array (F-USPA)}: A fixed sparse planar array with an inter-element spacing  of $20\lambda/3$.
    \item \textbf{Half-wavelength Uniform Planar Array (F-HwUPA)}: A conventional fixed planar array with a $\lambda/2$-element spacing.    
\end{itemize}

\begin{figure*}[t]
\centering
\subfloat[Ergodic sum rate for near-LoS case.] 
{\includegraphics[width=0.45\textwidth]{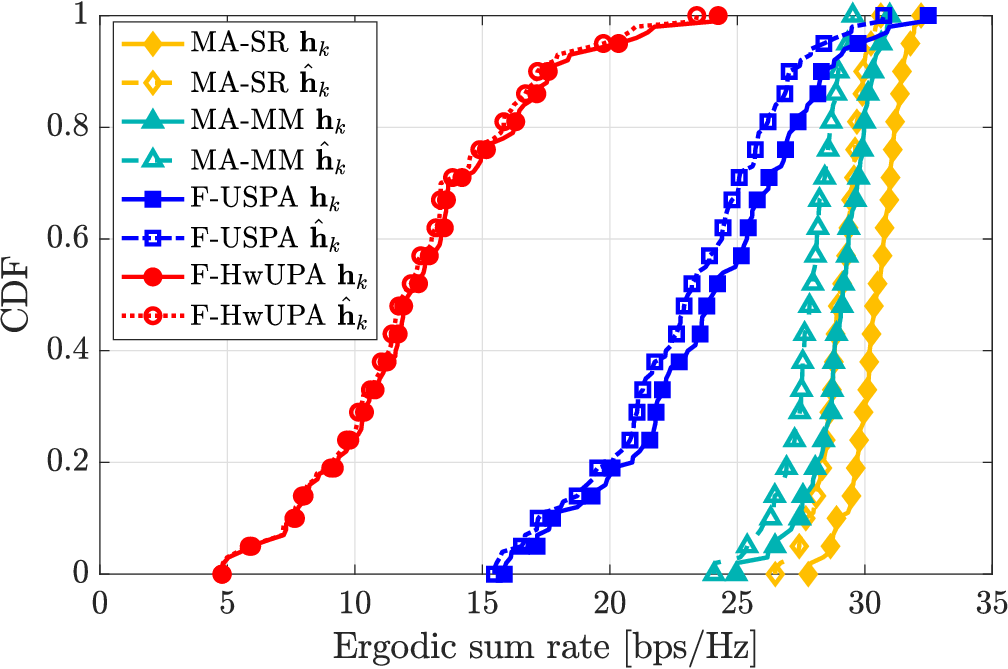}}\hspace{1cm}
\subfloat[Ergodic sum rate for rich-scattering case.]
{\includegraphics[width=0.45\textwidth]{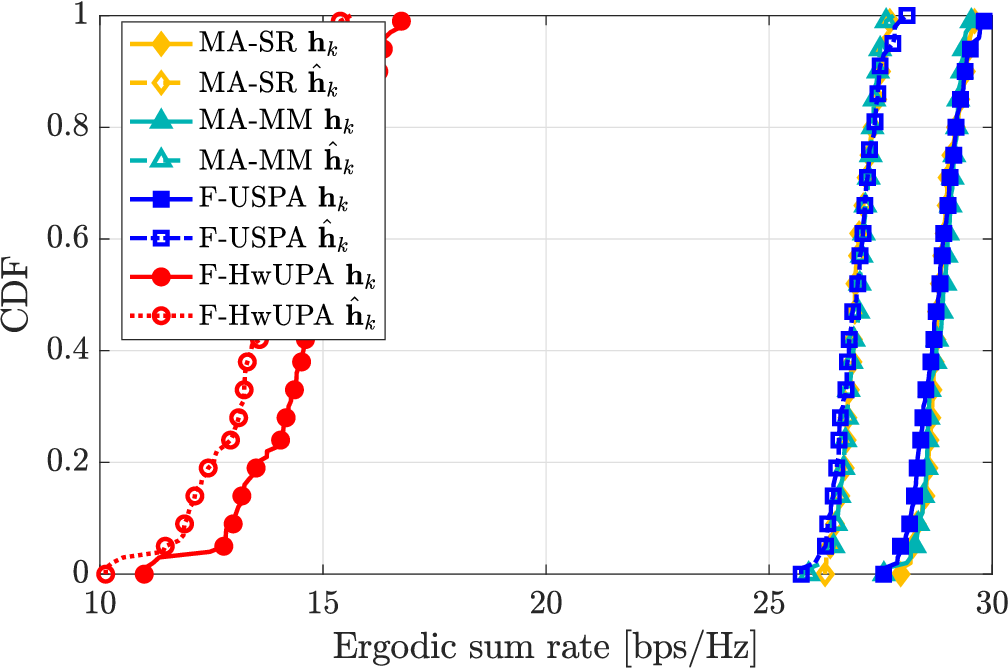}}
\centering
\caption{CDF of the sum rate for MA and fixed arrays under (a) sparse ($N_{\mathrm{cl}}=1$) and (b) rich ($N_{\mathrm{cl}}=10$) channels.}
\label{fig:cdf_ncl}
\end{figure*}

\subsection{Impact of Channel Structure}
\label{subsec:ncl}
To investigate how channel richness affects MA performance, we evaluate the system for varying $N_{\mathrm{cl}}$. Both MA configurations (MA-SR and MA-MM) are evaluated under a common ZF precoding scheme with waterfilling power allocation. The optimization follows the imperfect CSI model described in Section~\ref{sec:results}. Fig.~\ref{fig:cdf_ncl} shows CDFs of achievable ergodic sum rates for $N_{\mathrm{cl}} \in \{1, 10\}$ at SNR~$= 10$\,dB with $\tau_p = K = 10$.
For $N_{\mathrm{cl}} = 1$, the channel is near-LoS and highly directional. The low-rank (approximately rank-one) spatial covariance $\mathbf{R}_k$ gives strong spatial separation between users, and MA-PSO achieves the best performance: the genie-aided MA curve (solid yellow) is the highest, followed by MA with estimated channels (dashed yellow), then F-USPA (solid then dashed blue), and F-HwUPA performing the worst. The gap between dashed and solid curves is moderate, indicating that MMSE estimation remains accurate under the highly structured single-cluster channel. 
As $N_{\mathrm{cl}}$ increases to 10, the channel becomes richer in angular diversity and $\mathbf{R}_k$ increases in rank. User channels increasingly overlap in the angular space, reducing the spatial selectivity that MA repositioning exploits. Consequently, the CDF curves of MA and F-USPA, for both estimated and true channels, converge. Similar trends were observed for other user configurations (e.g., $K=5$), and are omitted for brevity. The key finding is that \emph{the MA advantage over fixed arrays is largest under sparse, near-LoS propagation and diminishes as scattering richness increases}.

\subsection{Cross-Objective Performance and the role of the precoder}

\begin{figure*}[t]
\centering
\subfloat[Minimum user rate with WF power allocation for both MA-SR and MA-MM.] 
{\includegraphics[width=0.45\textwidth]{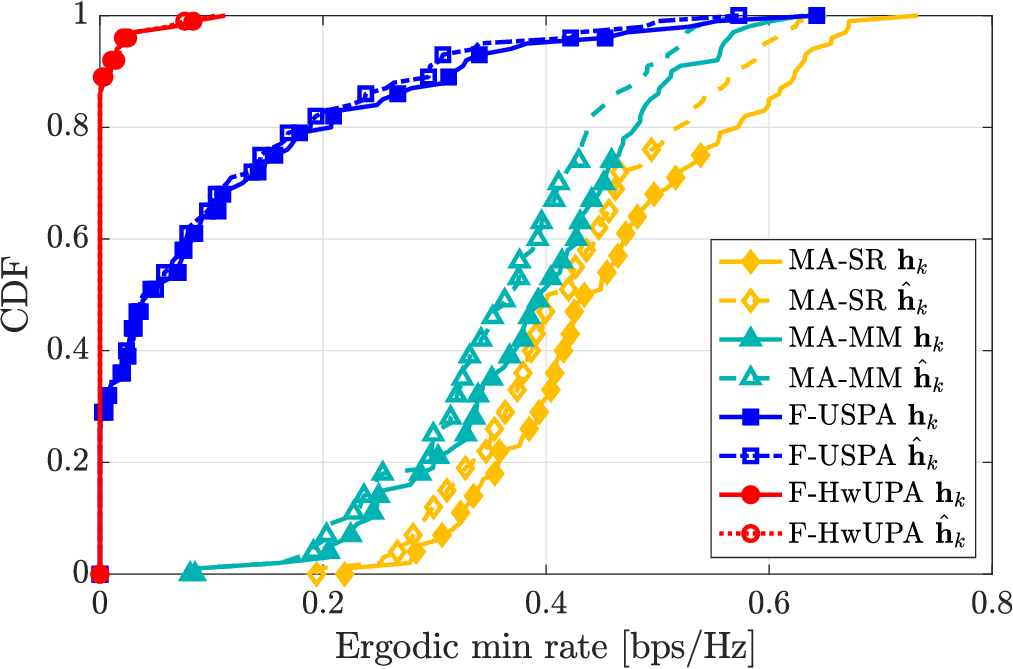}}\hspace{1cm}
\subfloat[Minimum user rate with equal power allocation for MA-MM.]
{\includegraphics[width=0.45\textwidth]{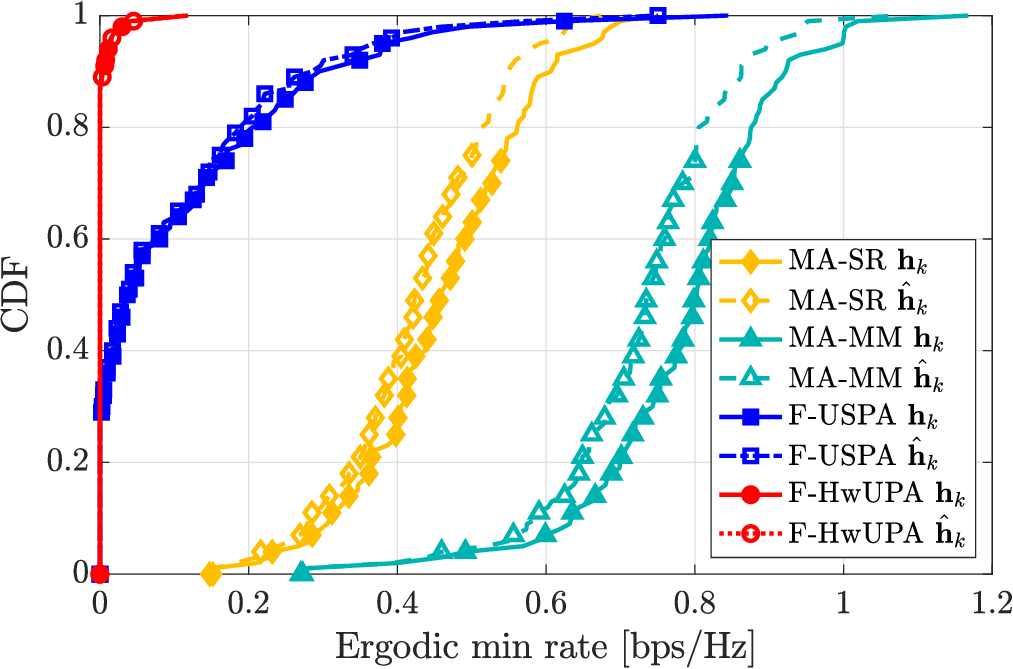}}
\centering
\caption{CDF of the minimum rate for MA-SR and MA-MM ($N_{\mathrm{cl}}=1$): (a) ZF with waterfilling and (b) ZF with equal power.}
\label{fig:cdf_pa}
\end{figure*}

Fig.~\ref{fig:cdf_pa}(a) shows the CDF of the minimum ergodic user rate for the same setup as in Section~\ref{subsec:ncl}, focusing on the sparse channel case ($N_{\mathrm{cl}}=1$). The two MA configurations differ only in the optimization objective used during PSO (sum-rate vs. max–min), while evaluation may use different precoders. Despite being optimized for fairness, MA-MM does not outperform MA-SR in terms of min-rate. This reveals that waterfilling actively counteracts the spatial fairness gains from the geometry: by allocating more power to users with stronger effective channels, the precoder suppresses the benefit of antenna positions that were found to geometrically favor weak users. In other words, the precoder dominates the geometry in this regime, and the geometric advantage of MA-MM vanishes when paired with a sum-rate-oriented power allocation.
In the second case, we pair MA-MM with equal-power ZF precoding, which does not penalize weak users through power allocation. To isolate the effect of the precoder, the antenna geometries obtained from PSO are kept fixed. Fig.~\ref{fig:cdf_pa}(b) shows that under this matched precoder, MA-MM achieves a clearly higher ergodic min-rate CDF than MA-SR, demonstrating that the geometry found by PSO-MaxMin genuinely encodes fairness information in the spatial domain. MA-SR retains its waterfilling precoder in this comparison, since it was optimized under that assumption.
These two cases together expose an important coupling between array geometry and precoder design: a fairness-oriented geometry only manifests its advantage when evaluated with a compatible precoder. This suggests that joint optimization of antenna placement and power allocation strategy is necessary to fully realize the benefits of movable antenna systems in multi-user scenarios. The key insight is that \emph{the performance of movable antenna systems is fundamentally governed by the interaction between array geometry and precoder design}. Geometry optimized for fairness provides gains only when paired with a compatible power allocation strategy, while mismatched designs can completely mask these benefits.

\section{Conclusion}\label{sec:conclude}

This paper presented a practical end-to-end framework for evaluating movable antenna systems under realistic imperfect CSI conditions. The framework encompasses uplink MMSE channel estimation, linear ZF precoding with waterfilling power allocation, and PSO-based antenna placement optimization driven solely by channel estimates. By reporting performance under both estimated and true channels for each antenna geometry, the framework directly quantifies the cost of CSI imperfection as a function of geometry and system parameters. The results demonstrate that the performance gains of movable antennas depend strongly on the propagation environment. In sparse, near-LoS channels with low angular spread, antenna repositioning yields substantial improvements over fixed deployments, whereas in rich-scattering environments these gains largely vanish as spatial selectivity decreases. A key finding concerns the coupling between array geometry and precoder design. When both MA configurations are evaluated under the same ZF precoder with waterfilling, the sum-rate-optimized array appears competitive with the fairness-optimized array even on min-rate metrics. However, this apparent robustness is a consequence of the waterfilling precoder masking the fairness information encoded in the geometry. When the fairness-oriented array is paired with a compatible equal-power precoder, it clearly outperforms the sum-rate array on the minimum ergodic user rate. This demonstrates that the benefits of a fairness-oriented antenna geometry are only realized when the precoder does not work against it; a precoder-geometry mismatch can completely mask the spatial advantage.
These findings highlight that movable antenna performance is scenario-dependent and that joint consideration of channel structure, optimization objective, and precoder design is essential for accurate system evaluation. Future work will investigate joint optimization of pilot sequences and antenna positions, extension to wideband frequency-selective channels, and adaptive positioning for time-varying multiple antenna user distributions.

\bibliographystyle{IEEEtran}
\bibliography{references}

\end{document}